\documentclass[twocolumn,showpacs,preprintnumbers,amsmath,amssymb]{revtex4}

\usepackage{dcolumn}
\usepackage{bm}
\usepackage{epsfig}
\usepackage{ulem}%
\usepackage{color}%

\begin{document}


\title{Magnetic flux density and the critical field in the intermediate state of type-I superconductors}

\author{V. Kozhevnikov$^{1}$, R. J. Wijngaarden$^2$, J. de Wit$^2$ and C. Van Haesendonck$^3$}

\affiliation{
$^1$Tulsa Community College, Tulsa, Oklahoma 74119, USA \\
$^2$VU University Amsterdam, 1081 HV Amsterdam, The Netherlands \\
$^3$Laboratory of Solid-State Physics and Magnetism, KU Leuven, BE-3001 Leuven, Belgium}. \\


\begin{abstract}
\noindent To address unsolved fundamental problems of the intermediate state (IS), the equilibrium magnetic flux structure and the critical
field in a high purity type-I superconductor (indium film) are investigated 
using magneto-optical imaging with a 3D vector magnet and electrical transport measurements. The least expected observation is that the
critical field in the IS can be as small as nearly 40\% of the thermodynamic critical field $H_c$. 
This indicates that the flux density in the \textit{bulk} of normal domains can be \textit{considerably} less than $H_c$, in apparent
contradiction with the long established paradigm, stating that the normal phase is unstable 
below $H_c$. Here we present a novel theoretical model consistently describing this and \textit{all} other properties of the IS. Moreover,
our model, based the rigorous thermodynamic treatment of observed laminar flux structure in a tilted field, 
allows for a \textit{quantitative} determination of the domain-wall parameter and the coherence length, and provides new insight into the
properties of all superconductors.

\end{abstract}\


\maketitle The interest in pattern formation in a big variety of physicochemical systems with spatially modulated phases \cite{Seul} has
sparked renewed attention to the intermediate state (IS) in type-I superconductors \cite{Ge, Peeters, Bending, Peeters_2, Prozorov_NP}, a
classical example of such systems with very rich physics \cite{Tinkham}.

Apart from the equilibrium flux pattern, unsolved fundamental problems of the IS include $B$ in the N-domains and the critical field
$H_{ci}$ for the IS-N transition. The IS provides access to one of the most fundamental parameters, namely the Pippard coherence length
$\xi_0$ (the size of Cooper pairs). However, a verified recipe to extract $\xi_0$ from the IS properties is missing. Similar problems, i.e.
$B$ in the vortex cores, or size of the cores, and extraction of microscopic parameters from properties of the mixed state (MS), are are
among the central problems of unconventional superconductivity \cite{Sonier_2011}. Like the IS, the MS is a two-phase superconducting state
with nonzero average flux density $\bar{B}$. Due to that the IS in type-I superconductors is ultimately related to the MS in type-II
materials \cite{Abrikosov}. In particular, an equilibrium spacing between normal (N) domains (those are vortices in type-II materials) in
both the IS and the MS is determined by long-range forces caused by inhomogeneity of the outside field near the surface \cite{Pearl,
Goldstein}. Therefore better understanding of the IS can provide new insights in properties of the MS.

Ever since Landau introduced a laminar model (LLM) for a slab in a perpendicular field \cite{Landau_37}, many models of the IS have been
proposed. However, none is fully adequate 
\cite{Prozorov}. Here we report on an experimental study of the IS performed with a high purity indium sample and introduce, for the first
time, a \textit{comprehensive} theoretical model for a slab in a tilted field. Our model is surprisingly simple. Nevertheless it allows for
\textit{quantitative} evaluation of the IS parameters, including $\xi_0$, and sheds new light on fundamental properties of \textit{all}
superconductors.

When a type-I superconductor with 
$\eta\neq0$ is subjected to a weak magnetic field $H$, the sample is in the Meissner state until $H$ reaches $H_i=(1-\eta)H_c$, $H_c$ being
the thermodynamic critical field \cite{Tinkham}. At $H_i$ the sample undergoes a transition to the IS, where it breaks up into N and
superconducting (S) domains with flux densities $B$ and zero, respectively. Under increasing $H$ the normal fraction $\rho_n= V_n/V$ ($V_n$
and $V$ are the volumes of N domains and of the sample, respectively) increases until the entire sample becomes normal at $H_{ci}$. In
accord with the standard paradigm stating that the N-phase is unstable at $B<H_c$ \cite{Gorter, Landafshitz_II}, $H_{ci}$ is assumed equal
to \cite{Landau_37} or slightly less than \cite{De Gennes, Tinkham} $H_c$. Below we show that this is true only for very thick samples.

The domain shape depends on many factors \cite{Huebener}. An important role is played by purity. Structural/chemical flaws reduce the
electron mean free path, hence increasing the Ginzburg-Landau (GL) parameter $\kappa$, and therefore decrease the S/N interface tension
$\gamma$, the latter being a ``driving force" in reaching the ground state. In addition, the flaws create pinning centers, hence reducing
domains' mobility. Therefore samples for studies of equilibrium flux patterns have to be pure and posses maximum possible $\gamma$ (minimum
$\kappa$) \cite{Faber}. The equilibrium flux pattern is well established for a cylinder in a perpendicular field and a slab in a strongly
tilted field \cite{Abrikosov}. For the latter $\eta=1$ ($H_i=0$) and domains are ordered laminae. The IS in the slab can be investigated
using magneto-optics (MO) \cite{Huebener}. This is the experiment we performed.

We focus on the following questions. (1) How does the flux density $B$ and the critical field $H_{ci}$ depend on the material parameters,
and how do these quantities evolve with magnitude and orientation of $H$? (2) How can the domain-wall parameter $\delta$ (and therefore
also the GL $\xi(T)$ and the Pippard $\xi_0$ coherence lengths) be inferred from the laminar structure? The relationship between $\delta$,
$\xi(T)$ and $\xi_0$ depends on the material and its purity. For the pure-limit Pippard superconductors ($\kappa\ll1$)
$\delta(T)=1.89\xi(T)=1.4\xi_0/(1-t)^{0.5}$, where $t=T/T_c$ \cite{Tinkham}.

At first sight the answers are known \cite{Landafshitz_II, De Gennes, Tinkham, Abrikosov}. However, $\xi_0$ in Sn and In, determined from
the field profile measurements (310 nm in Sn and 380 nm in In) \cite{VK}, differ from $\xi_0$ calculated from $\delta(0)$ obtained from the
IS (180~\cite{Sharvin-Sn} and 240~nm~\cite{Sharvin-In}, respectively). Since in both cases the samples were very pure, this signals the
inadequacy of models used either in Ref.~\onlinecite{VK} or in Refs.~\onlinecite{Sharvin-Sn, Sharvin-In}. Resolving this contradiction was
the original motivation for this work.

The IS structure was first treated by Landau in 1937 \cite{Landau_37}. He established the concept of the surface tension $\gamma$ ($\delta$
was later defined as $8\pi\gamma/H_c^2$ \cite{BM}) and proposed the LLM. Assuming that $B = H_c$, Landau calculated the shape of rounded
corners of a cross section of the S-laminae. The rounded corners yield an excess energy of the system competing with the interface energy.
Minimizing the sum of these energy contributions, Landau obtained the period of the laminar structure $D=\sqrt{\delta d/f_L(\rho_n)}$,
where $d$ is the sample thickness and $f_L(\rho_n)$ is the ``spacing function" with $\rho_n=h\equiv H/H_c$; $f_L$ is determined by the
shape of the corners \cite{Landafshitz_II}.

Soon thereafter Landau admitted that the LLM is unstable because $B < H_c$ near the surface. However, a ``branching" model \cite{BM},
proposed instead of the LLM, was disproved by Meshkovsky and Shalnikov in 1947. Owing to that in 1951 Lifshitz and Sharvin turned back to
the LLM and calculated $f_L(\rho_n)$ and $B$ numerically. Identical results have later been obtained analytically \cite{Fortini}. In the
LLM $B$, being $H_c$ in the bulk, is $0.66H_c$ near the surface at low $H$ and increases up to $H_c$ at the IS-N transition. Fifty years
later direct bulk $\mu$SR measurements of $B$ in a Sn slab in a perpendicular field revealed that $B(H)$ starts from $H_c$ and decreases
with $H$ down to $H_{ci}$ \cite{Egorov}. Such a dependence for $B(H)$ had been anticipated by Tinkham \cite{Tinkham}.

De Gennes  \cite{De Gennes} noticed that a positive $\gamma$ should reduce $H_{ci}$. Assuming a small reduction, de Gennes obtained for the
transverse configuration $H_{ci\perp}= H_c[1-0.9(\delta/d)^{0.5}]$.

Tinkham \cite{Tinkham} recognized that the dominant contribution to the excess energy term comes from inhomogeneity of the field outside
the sample and therefore roundness of the corners can be neglected. Tinkham computed this energy by introducing a ``healing length" $L_h$
over which the field relaxes to its uniform state: $L_h^{-1}=D_n^{-1}+D_s^{-1}$, where $D_n$ and $D_s$ are the width of the S- and
N-laminae, respectively. Supposing a rectangular cross section of the laminae and hence a uniform $B$, allowed to be somewhat less than
$H_c$, Tinkham obtained
\begin{eqnarray}
H_{ci\perp}=H_c[(1+4\delta/d)^{0.5}-2(\delta/d)^{0.5}].\label{eq:Tinkham}
\end{eqnarray}

The structure expected from the LLM has \textit{never} been observed. Images reported from the 1950s onwards revealed intricate flux
patterns, often forming corrugated laminae \cite{Huebener}.

Sharvin \cite{Sharvin-Sn} was the first to observe the ordered laminar pattern in a strongly tilted field for 2-mm thick Sn
\cite{Sharvin-Sn} and In \cite{Sharvin-In} samples. He measured $D$ at different temperatures and calculated $\delta(T)$ using an extended
LLM assuming $B_\parallel=H_\parallel$ ($B_\parallel$ and $H_\parallel$ are in-plane components of $B$ and $H$, respectively) on top of
Landau's original assumption that $B=H_c$. Sharvin's equation for $D$ is
\begin{eqnarray}
D^2=\frac{\delta
d}{f_L(\rho_n)}\cdot\frac{H_c^2}{H_c^2-H_\parallel^2}.\label{eq:Sharvin}
\end{eqnarray}

The values of $\delta(0)$ obtained for In and Sn using Eq.\,(\ref{eq:Sharvin}) \cite{Sharvin-In, Sharvin-Sn} are those from which we
started our story. Faber \cite{Faber} criticized Eq.\,(\ref{eq:Sharvin}) arguing that $H_\parallel$ can alter the shape of the corners. One
may add that if $B_\|=H_\|$ the magnetic flux is not conserved and therefore the energy balance in the system must be reconsidered.

From the above it follows that the  extended LLM \cite{Sharvin-Sn} is questionable. Therefore the values of 
$\delta(0)$ obtained in Refs.\,\onlinecite{Sharvin-In, Sharvin-Sn} are questionable as well. However, the $\delta(T)$ dependence obtained
by Sharvin is correct because it agrees with the GL theory. (Historically it was in reverse order \cite{Tinkham}.) Besides, the question of
how to extract $\delta$ from the IS pattern in a tilted field remains to be answered. The latter two issues, along with the questions on
$B$ and $H_{ci}$, are addressed below.

\begin{figure}
\epsfig{file=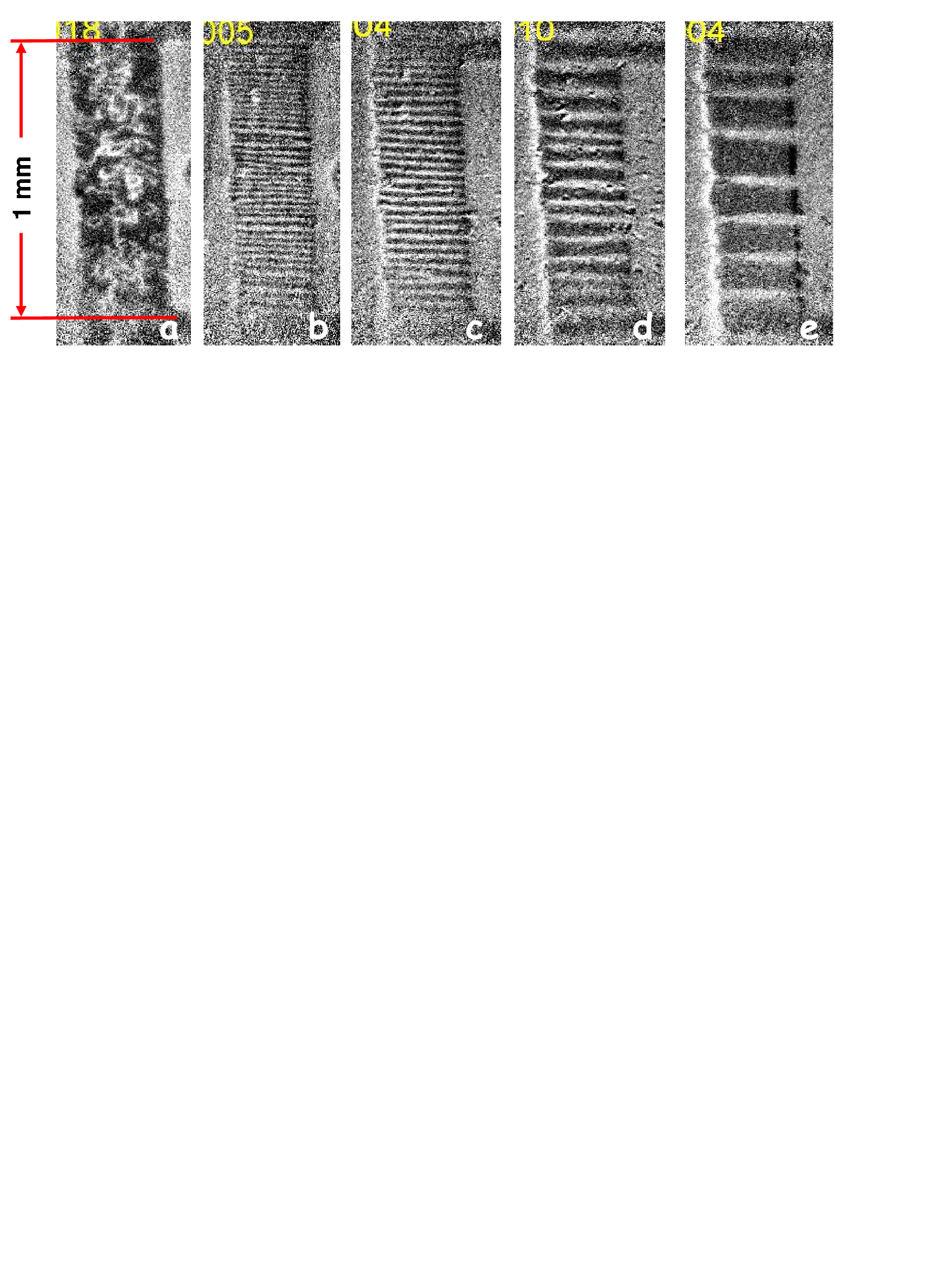,width=8 cm}
\caption{\label{fig:epsart} MO images taken at 2.5 K.
Superconducting regions are black. ($H_\parallel$, $H_\perp$; in
Oe): a, (0, 1); b, (60, 8); c, (100, 6); d, (110, 3); e, (115, 1.3).
}
\end{figure}
The MO imaging was achieved using a set-up equipped with a 3D vector magnet \cite{Rinke2}. The sample was a 2.5 $\mu$m thick In film 
on a SiO$_2$ wafer. The film residual resistivity ratio was 540. The other film characteristics were the same as 
in Ref.\,\onlinecite{VK}. Overall, the film is a Pippard superconductor ($\kappa\simeq0.07$) in the pure limit. The sample length was 1 mm
and the ratio width/thickness was 120, implying that for the perpendicular and parallel fields $\eta$ is $1$ and $0$, respectively. Images
were taken simultaneously with measurements of the electrical resistance $R$ using a small low-frequency (11\textbf\,Hz) AC current.

Typical images are presented in Fig.\,1. The flux patterns are laminae independent of the history of the applied field. At $H_\perp
\lesssim 1$ Oe fractionated laminae were seen in some runs. The laminae are planar and ordered at $H_\parallel \gtrsim 0.5 H_c$. At smaller
$H_\parallel$ ($= 0.47 H_c$ at 1.67 K) slight wave-like corrugations appear. In a perpendicular field  the laminae are disordered
(Fig.\,1a). This suggests that the laminar structure is the ground state topology of the IS. The same conclusion was drawn by
Faber~\cite{Faber}.

While $D_n$ ($D_s$) increases (decreases) and $R$ varies linearly  with $H_\perp$, the period $D=D_n+D_s$, being dependent on
$H_\parallel$, is constant for $0.3\lesssim H_\perp/H_{ic\perp} \lesssim 0.8$. Near $H_\perp=0$ and $H_{ci\perp}$ the number of laminae
decreases. The IS-N transition for decreasing field is accompanied by deep supercooling of the N-state. This confirms the high purity of
the sample and verifies that the IS-N transition is a first order phase transition \cite{Abrikosov}.

Figure 2 presents the sample phase diagram $H_c(T)=H_{c\parallel}(T)$ measured with a DC magnetometer and used for \textit{in situ}
temperature determination. 
It is compared with data measured on other samples and the data from Ref.\,\onlinecite{Finnemore}. The lower curve presents
$H_{ci\perp}(T)$ at $H_\parallel =0$, determined from disappearance of the last S-lamina in the images and from $R(H_\perp)$ measurements;
the two perfectly coincide. As seen, $H_{ci\perp}(T)$ is \textit{less than half} of $H_c(T)$. The stars represent $H_{ci\perp}$ calculated
from Eq.\,(\ref{eq:Tinkham}) with $\xi_0$ = 380\,nm and are clearly consistent with the experimental data. De Gennes' formula yields
$H_{ci\perp}$ considerably exceeding the experimental data. Hence we use Tinkham's interpretation for the excess energy term.

\textbf{The model.} Since we adopt Tinkham's approach and \textit{in agreement with the experimental images}, the N- and S-domains are
assumed to be rectangular parallelepipeds extending in the $H_\parallel$ direction. The contribution of the negative surface tension at the
S/vacuum interface is neglected since the penetration depth is much less than the sample thickness.  Magnetostriction effects are neglected
as well. The out-of-plane and in-plane demagnetizing factors are $\eta_\perp=1$ and $\eta_\parallel=0$, respectively. The former means that
$B_\perp\rho_n=H_\perp$ (conservation of the flux of the out-of-plane component of the magnetic field), whereas the latter means that
$B_\parallel=H_\parallel$ and therefore the flux of $B_\parallel$ is not conserved. Hence, the appropriate thermodynamic potential is
$\tilde{F}\equiv\tilde{f}V = F - V(B_\parallel H_\parallel/4\pi) = F - V_n(H_\parallel^2/4\pi)$, where $F$ is the free energy and the
second term accounts for the work done by the generator to maintain $H_\parallel$ \cite{Landafshitz_II}. This term is the \textit{key
distinctive element} of our model. We note that this term is neither small (it can exceed the condensation energy) nor trivial (its
omission or 
incorrect form leads to violation of the limiting cases).

Summing the sample free energy at zero field $V[f_{n0}-H_c^2(1-\rho_n)/8\pi)]$, the energy of the field $B$ in the sample
$V\rho_n(B_\perp^2+B_\parallel^2)/8\pi$, the energy of the S/N interfaces $V2\gamma/D$ and the excess energy of the field over the healing
length $2VL_h(\rho_nB_\perp^2-H_\perp^2)/8\pi d$, one obtains
\begin{multline}
\tilde{f}=f_{n0}-(1-\rho_n)\frac{H_c^2}{8\pi}+\frac{H_\perp^2}{8\pi\rho_n}-\rho_n\frac{H_\parallel^2}{8\pi}
+\\2\frac{H_c^2}{8\pi}\frac{\delta}{D}+2\frac{H_\perp^2}{8\pi}\frac{D}{d}(1-\rho_n)^2,\label{eq:g}
\end{multline}
where $f_{n0}$ is the free energy density of the sample in the
normal state at zero field.

Similar to the LLM, competition between the last two terms provides the equilibrium $D$:
\begin{eqnarray}
D^2=\frac{d\delta}{\rho_n^2(1-\rho_n)^2}\frac{H_c^2}{B_\perp^2}=\frac{d\delta}{(1-\rho_n)^2}\frac{H_c^2}{H_\perp^2}.\label{eq:D}
\end{eqnarray}

Substituting (\ref{eq:D}) in (\ref{eq:g}) and then minimizing $\tilde{f}(\rho_n)$ one obtains the equilibrium $\rho_n$:
\begin{eqnarray}
\rho_n^2=h_\perp^2/(1-4h_\perp\sqrt{\delta/d}-h_\parallel^2).\label{eq:rho_n}
\end{eqnarray}
where $h_\perp=H_\perp/H_c$ and $h_\parallel=H_\parallel/H_c$.

At the IS-N transition $\rho_n$=1, therefore
\begin{eqnarray}
h_{ci\perp}=\sqrt{4(\delta/d)+1-h_\parallel^2}-2\sqrt{\delta/d}.\label{eq:hc-per}
\end{eqnarray}

Finally, the reduced flux density $b=B/H_c$ is
\begin{eqnarray}
b^2=b_\perp^2+b_\parallel^2=h_\perp^2/\rho_n^2+h_\parallel^2=1-4h_\perp\sqrt{\delta/d}.\label{eq:b}
\end{eqnarray}

The model satisfies the limiting cases, i.e., $\rho_n\rightarrow0$ at $H_\perp\rightarrow0$ and $H_{ci\perp}\rightarrow0$ at
$H_\parallel\rightarrow H_c$. In very thick samples ($\sqrt{\delta/d}\ll1$) $B=H_c$ and Eq.\,(\ref{eq:D}) converts to
Eq.\,(\ref{eq:Sharvin}) if $\rho_n^2(1-\rho_n)^2$ is replaced by $f_L(\rho_n)$; this explains the correctness of the temperature
dependence $\delta(T)$ in Refs.\,\onlinecite{Sharvin-Sn, Sharvin-In}.

For a perpendicular field ($H_\parallel=0$) we have that (a) according to Eq.\,(\ref{eq:b}) $B$ \textit{decreases} with increasing $H$ from
$H_c$ down to $H_{ci}$, in agreement with the $\mu$SR results \cite{Egorov}; and that (b) Eq.\,(\ref{eq:hc-per}) reduces to
Eq.\,(\ref{eq:Tinkham}), implying that the theoretical points in Fig.\,2 are the same in our model. Hence, our model, developed for a
regular laminar structure, can be used for irregular laminar patterns as well.

\begin{figure}
\epsfig{file=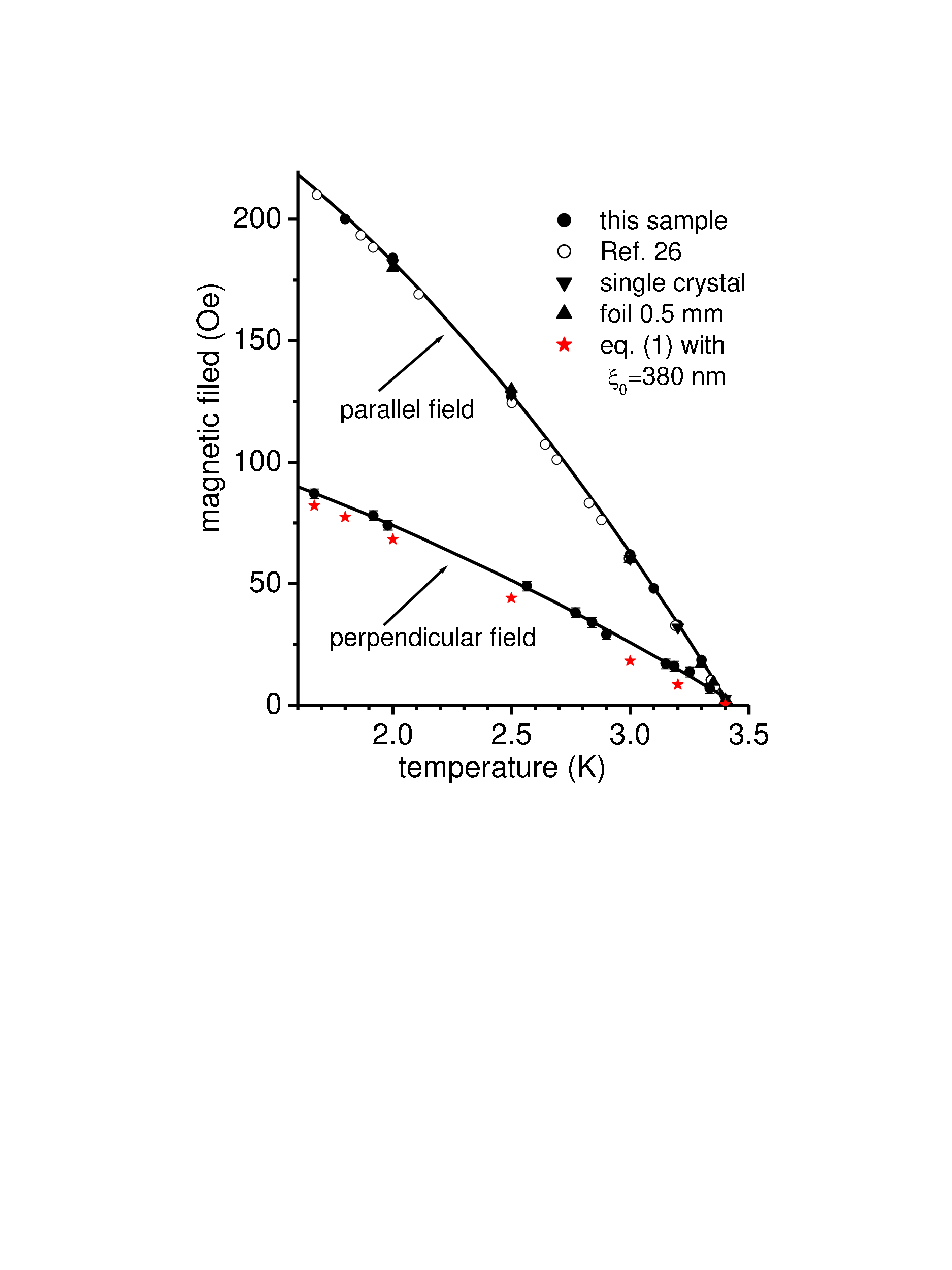,width=5.5 cm}
\caption{\label{fig:epsart}The phase diagram of our sample in
parallel and perpendicular fields. The full curves are parabolic
fits of the data obtained with our sample.}
\end{figure}

In Fig.\,3 the data for $H_{ci\perp}$ at nonzero $H_\parallel$ are
compared to the $H_{ci\perp}(H_\parallel)$ dependence given by
Eq.\,(\ref{eq:hc-per}). We find that Eq.\,(\ref{eq:hc-per})
correctly describes the experimental data.
\begin{figure}
\epsfig{file=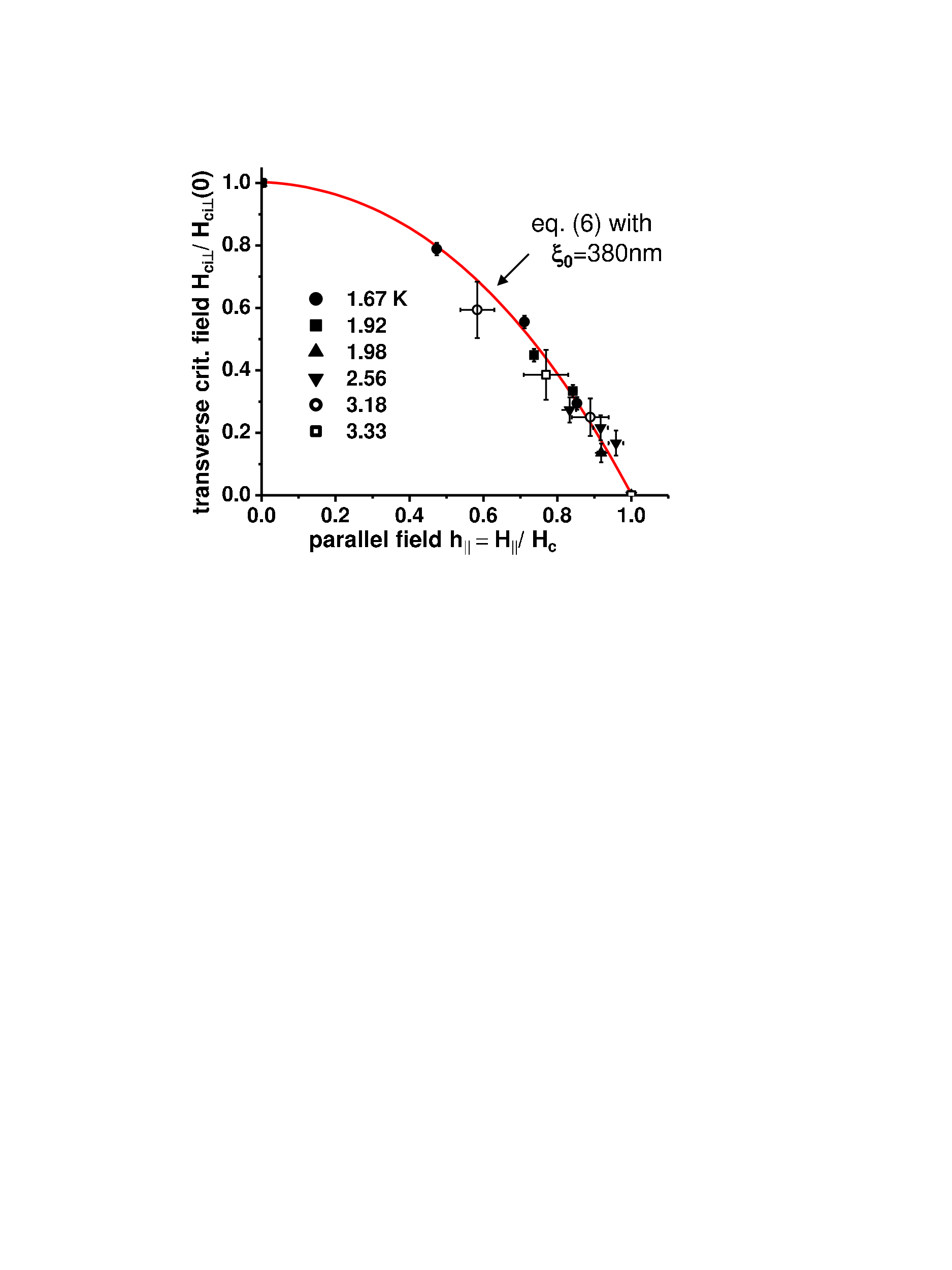,width=5.0 cm}
\caption{\label{fig:epsart} The perpendicular critical field
$H_{ci\perp}/H_{ci\perp}(0)$ versus $h_\parallel=H_\parallel/H_c$.
$H_{ci\perp}(0)$ is $H_{ci\perp}$ at $H_\parallel=0$.}
\end{figure}
\begin{figure}
\epsfig{file=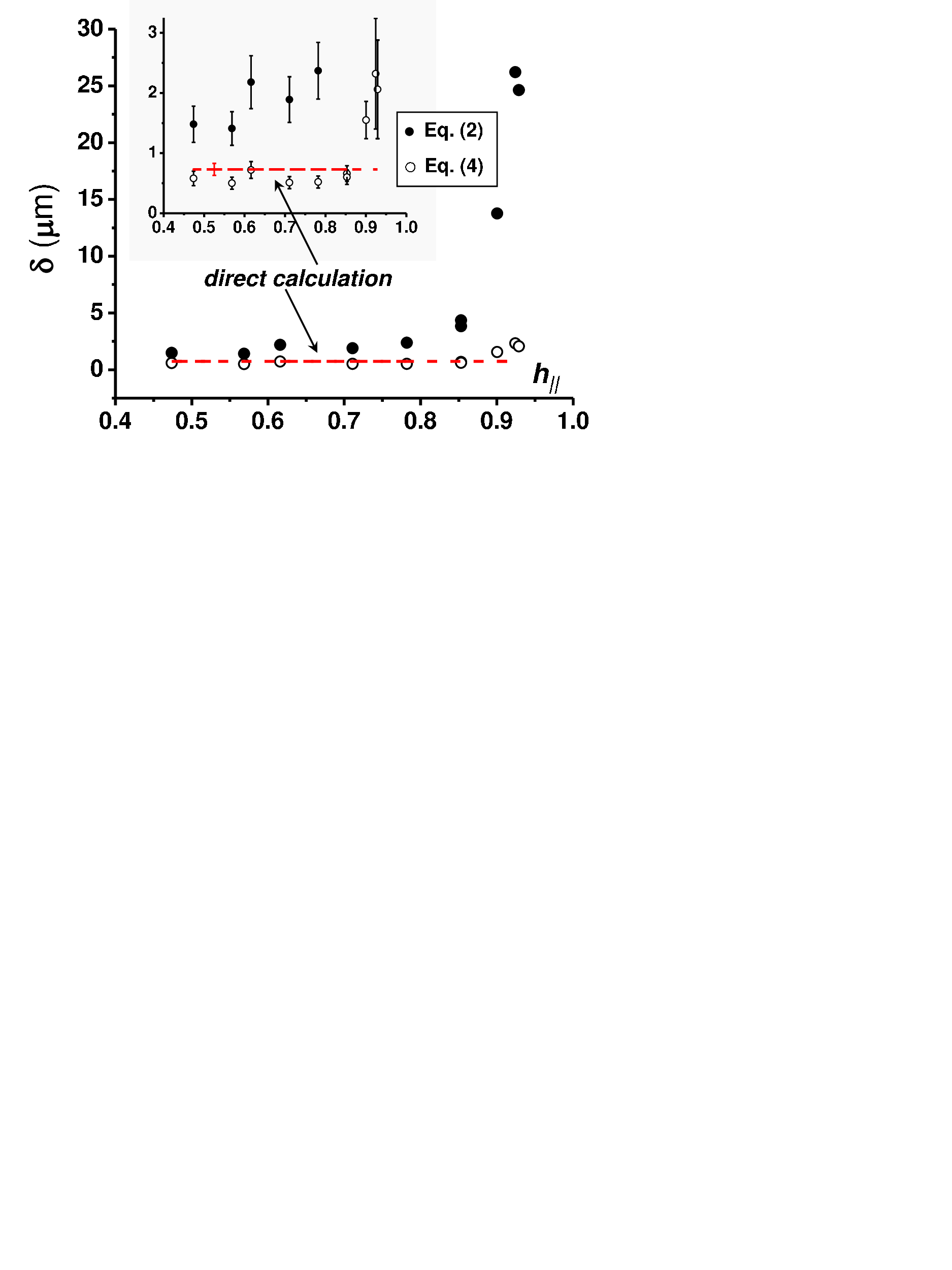,width=5.2 cm}
\caption{\label{fig:epsart} The domain-wall parameter $\delta$
inferred from MO images taken at $T$\,=\,1.7\,K using
Eq.\,(\ref{eq:Sharvin}) and Eq.\,(\ref{eq:D}) plotted on two scales.
The dashed line represents the directly calculated
$\delta=1.4\xi_0/(1-T/T_c)^{0.5}$, $T_c=3.415$\,K and
$\xi_0=380$\,nm.}
\end{figure}

Figure 4 presents the results for $\delta$ at $T$\,=\,1.7\,K obtained in three ways. The circles represent $\delta$ calculated from
Eq.\,(\ref{eq:Sharvin}) and Eq.\,(\ref{eq:D}) using our experimental data for the average $D$ and $H_\perp$ at $\rho_n=0.5$. The dashed
line in Fig.\,4 represents $\delta$ directly calculated using $\xi_0$ = 380 nm. We find that the results following from Eq.\,(\ref{eq:D})
agree with the directly calculated value of $\delta$. However this is not the case for $\delta$ obtained using Eq.\,(\ref{eq:Sharvin}).

Figure 5 presents data for the average $D$ obtained in two runs at $T=1.7\,K$ and $H_\parallel=0.85H_c$ and corresponding theoretical
curves following from Eqs.\,(\ref{eq:Sharvin}) and (\ref{eq:D}). In Eq.\,(\ref{eq:Sharvin}) $D$ is controlled by $f_L$; in our model the
spacing function is $f=(1-\rho_n)^2 h_\perp^2$ with $\rho_n$ and $h_\perp$ linked by Eq.\,(\ref{eq:rho_n}). We find that both $f_L$ and $f$
qualitatively reproduce the experimental data. However $f_L$ ``works'' better at high values of $H_\perp/H_{ci\perp}$, whereas $f$ is
better at small $H_\perp$. This indicates the importance of the rounded corners at high $H_\perp$, where inhomogeneity of $H_\perp$ is
minimal.

\begin{figure}
\epsfig{file=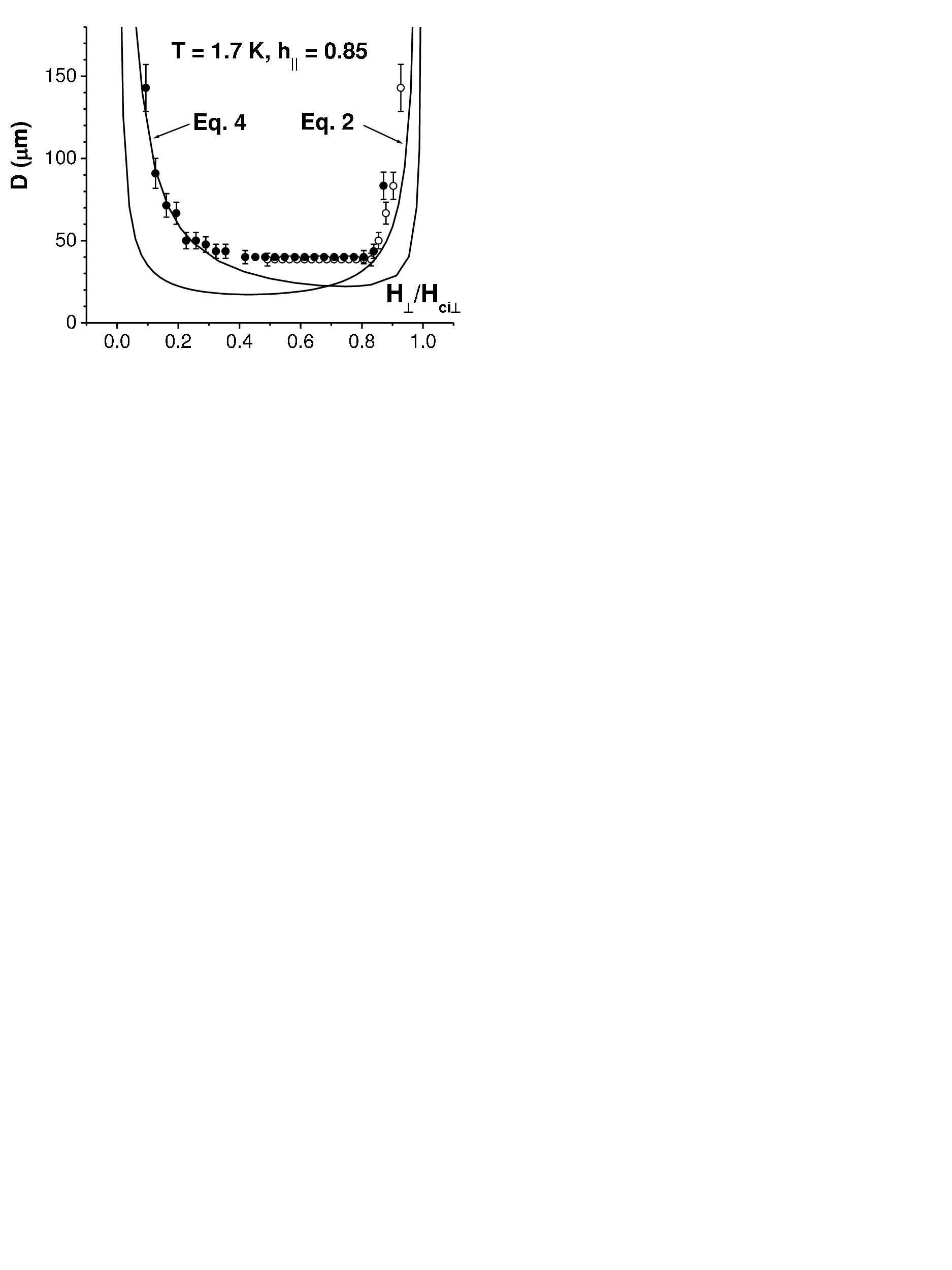,width=5. cm}
\caption{\label{fig:epsart} Experimental data (circles) and
theoretical curves for the period of the laminar structure. Solid
circles are the data obtained at increasing $H_\perp$ with the
sample cooled at $H_\perp=0$. Open circles are the data taken after
$H_\perp$ was reduced from above $H_{ci\perp}$ down to
$0.5H_{ci\perp}$.}
\end{figure}

\textbf{In summary}, (i) we have performed a magneto-optical study of the IS in the high purity type-I superconductor resulting in a novel
and comprehensive model of the IS for a slab in a tilted magnetic field, which includes the perpendicular field as the limiting case. The
model is a \textit{good first order approximation} of the IS in a slab as advanced by Landau many years ago. (ii) We have shown that a
superconducting system in search for the lowest free energy may opt to keep $B$ in the bulk of N-domains \textit{considerably} smaller than
$H_c$. This alters the paradigm stating that this is impossible. In type-II superconductors the energy of the field $B$ in the vortex cores
and the energy of the inhomogeneous field near the surface are composite parts of a sample free energy in the MS. Therefore variation of
$B$ in the core and its dependents on the sample thickness can be expected in these materials as well. In particular, this can be a factor
responsible for the observed dependence of the vortex core size on the applied field \cite{Sonier}.

A weak point of our model is the oversimplified form of $L_h$ and neglect of the effect of the rounded corners. This is the main reason for
the discrepancy between the experimental data and the theoretical curve in Fig.\,5 and the deviation of $\rho_n(h_\perp)$ in Eq.\,(6) from
the linear dependence following from the linearity of $R(H_\perp)$. To resolve this issue measurements of the magnetic field near the
surface outside and inside the sample are required.

This work was supported by NSF (DMR 0904157) and by the Research Foundation -- Flanders (FWO).

\end{document}